# A Variation Evolving Method for Optimal Control

Sheng ZHANG, En-Mi YONG, Wei-Qi QIAN, and Kai-Feng HE

(2017.01)

*Abstract:* A new method for the optimal solutions is proposed. Originating from the continuous-time dynamics stability theory in the control field, the optimal solution is anticipated to be obtained in an asymptotically evolving way. By introducing a virtual dimension——the variation time, a dynamic system that describes the variation motion is deduced from the Optimal Control Problem (OCP), and the optimal solution is its equilibrium point. Through this method, the intractable OCP is transformed to the Initial-value Problem (IVP) and it may be solved with mature Ordinary Differential Equation (ODE) numerical integration methods. Especially, the deduced dynamic system is globally stable, so any initial value will evolve to the extremal solution ultimately.

## I. Introduction

Optimal control theory aims to determine the inputs to a dynamic system that optimize a specified performance index while satisfying constraints on the motion of the system. It is closely related to engineering and has been widely studied [1]. Because of the complexity, usually Optimal Control Problems (OCPs) are solved with numerical methods. Various numerical methods are developed and generally they are divided into two classes, namely, the direct methods and the indirect methods [2]. The direct methods discretize the control or/and state variables to obtain the Nonlinear Programming (NLP) problem, for example, the widely-used direct shooting method [2] and the classic collocation method [3]. These methods are easy to apply, whereas the results obtained are usually suboptimal [4], and the optimal may be infinitely approached. The indirect methods transform the OCP to a boundary-value problem (BVP) through the optimality conditions. Typical methods of this type include the well-known indirect shooting method [2] and the novel symplectic method [5]. Although be more precise, the indirect methods often suffer from the significant numerical difficulty due to ill-conditioning of the Hamiltonian dynamics, that is, the stability of costates dynamics is adverse to that of the state dynamics [6]. The recent development, representatively the Pseudo-spectral (PS) method [7], blends the two types of methods, as it unifies the NLP and the BVP in a dualization view [8]. Such methods inherit the advantages of both types and blur their difference.

The authors are with the Computational Aerodynamics Institution, China Aerodynamics Research and Development Center, Mianyang, 621000, China. (e-mail: zszhangshengzs@hotmail.com).

Theories in the control field often enlighten strategies for the optimal control computation, for example, the nonlinear variable transformation to reduce the variables [9]. In this paper, the dynamics stability theory [10], regarding the continuous-time system evolving, motivates the new method──the Variation Evolving Method (VEM), and the OCPs are transformed to the Initial-value Problems (IVPs). A virtual variation time dimension $\tau$, distinguished from the normal time variable $t$ in the OCP, is introduced to describe the evolution to the optimal solution. Generally, the BVP can be transformed to the IVP only for special cases (see [11] and [12] for example), while here typical OCPs may be transformed to the IVPs with respect to $\tau$, from a new angle of view. By guaranteeing the extremum as the equilibrium point of the deduced dynamic system, the optimal solution will be gradually approached. Fig. 1 illustrates the idea of the VEM in solving the OCP. Through variation motion, the initial guess of variable will evolve to the optimal solution.

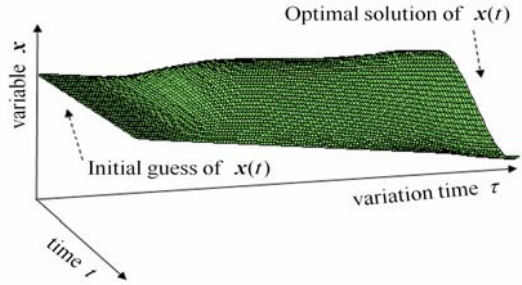

Fig. 1. The illustration of the variable evolving in the VEM.

In the following, first the foundational VEM is presented by applying to the unconstrained calculus-of-variations problem. Then the optimal control for the dynamic system is solved. We present some comments about the work in Section IV and Section V concludes the paper at the end.

## II. THE FOUNDATIONAL VARIATION EVOLVING METHOD

In the paper, our work is built upon the assumption that the solution for the optimization problem exists. We do not describe the existing conditions for the purpose of brevity. Relevant researches such as the Filippov-Cesari theorem are documented in [13]. From this premise, we demonstrate the foundational VEM by solving the unconstrained calculus-of-variations problem that is defined as

**Problem 1:** For the following functional depending on variable vector $y(t) \in \mathbb{R}^n$

$$J = \int_{t_0}^{t_f} F(y(t), \dot{y}(t), t) \, dt \tag{1}$$

where $t \in \mathbb{R}$ is time. The elements of $y$ belong to $C^2[t_0, t_f]$, which denotes the set of variables with continuous second-order derivatives (indicated by the superscript). The function $F: \mathbb{R}^n \times \mathbb{R}^n \times \mathbb{R} \to \mathbb{R}$ and its first-order and second-order partial derivatives are continuous with respect to $y$, its time derivative $\dot{y} = \dfrac{\mathrm{d}y}{\mathrm{d}t}$ and $t$. $t_0$ and $t_f$ are the fixed initial and terminal time. Find the extremum $\hat{y}$ that minimizes $J$, i.e.

$$\hat{y} = \arg\min(J) \tag{2}$$

Through the variation theory, the extremum for this functional satisfies the Euler-Lagrange equation [14]

$$F_y - \frac{\mathrm{d}}{\mathrm{d}t}(F_{\dot{y}}) = \mathbf{0} \tag{3}$$

and the boundary conditions

$$F_{\dot{y}}(t_0) = \mathbf{0} \tag{4}$$

$$F_{\dot{y}}(t_f) = \mathbf{0} \tag{5}$$

where $F_y = \dfrac{\partial F}{\partial y}$ and $F_{\dot{y}} = \dfrac{\partial F}{\partial \dot{y}}$ are the shorthand notations of partial derivatives. Instead of directly solving the BVP defined by Eqs. (3)-(5) to get the solution, it is enlightened by the states evolving in continuous-time stable dynamic systems that any initial guess $\tilde{y}(t)$, whose elements belong to $C^2[t_0, t_f]$, will evolve to the extremum along the variation dimension. Like the decrease of a Lyapunov function, through introducing the variation time, $\tau$, if $J$ decreases with respect to $\tau$, i.e., $\dfrac{\delta J}{\delta \tau} < 0$, we may finally obtain the optimal solution. Differentiating Eq. (1) with respect to $\tau$ produces

$$\frac{\delta J}{\delta \tau} = F_{\dot{y}}^{\mathrm{T}} \frac{\delta y}{\delta \tau}\bigg|_{t_f} - F_{\dot{y}}^{\mathrm{T}} \frac{\delta y}{\delta \tau}\bigg|_{t_0} + \int_{t_0}^{t_f} \left(\left[F_y - \frac{\mathrm{d}}{\mathrm{d}t}(F_{\dot{y}})\right]^{\mathrm{T}} \frac{\delta y}{\delta \tau}\right) \mathrm{d}t \tag{6}$$

where the superscript "T" denotes the transpose operator. By enforcing $\dfrac{\delta J}{\delta \tau} \leq 0$, we may set that

$$\frac{\delta y}{\delta \tau} = -K\left(F_y - \frac{\mathrm{d}}{\mathrm{d}t}(F_{\dot{y}})\right), \quad t \in (t_0, t_f) \tag{7}$$

$$\frac{\delta y(t_0)}{\delta \tau} = KF_{\dot{y}}(t_0) \tag{8}$$

$$\frac{\delta y(t_f)}{\delta \tau} = -KF_{\dot{y}}(t_f) \tag{9}$$

where $\boldsymbol{K} = \mathrm{diag}(k_1, k_2, ..., k_n)$ is a positive diagonal matrix. To solve the variation dynamic equations, the initial guess $\boldsymbol{y}(t)\big|_{\tau=0} = \tilde{\boldsymbol{y}}(t)$, $t \in [t_0, t_f]$ may be arbitrary variable profile with elements belonging to $C^2[t_0, t_f]$. Equations (7)-(9) describe the variation motion of $\boldsymbol{y}(t)$ starting from $\tilde{\boldsymbol{y}}(t)$, and the motion is directed to the extremum.

**Theorem 1:** Solving the IVP defined by Eqs. (7)-(9) from the VEM, when $\tau \to +\infty$, $\boldsymbol{y}(t)$ will satisfy the optimality conditions of Problem 1.

Proof: Substituting Eqs. (7)-(9) to Eq. (6), we have $\dfrac{\delta J}{\delta \tau} \leq 0$. The functional $J$ will decrease until $\dfrac{\delta J}{\delta \tau} = 0$, which occurs when $\tau \to +\infty$ due to the asymptotical approach. When $\dfrac{\delta J}{\delta \tau} = 0$, this determines the optimal conditions, namely, Eqs. (3)-(5). ∎

**Remark 1:** If in Problem 1 the boundary conditions of the variable vector $\boldsymbol{y}(t)$ are prescribed as $\boldsymbol{y}(t_0) = \boldsymbol{y}_0$ and $\boldsymbol{y}(t_f) = \boldsymbol{y}_f$, then we may solve the problem using Eq. (7) and

$$\frac{\delta \boldsymbol{y}(t_0)}{\delta \tau} = \boldsymbol{0} \tag{10}$$

$$\frac{\delta \boldsymbol{y}(t_f)}{\delta \tau} = \boldsymbol{0} \tag{11}$$

with the initial guess $\boldsymbol{y}(t_0)\big|_{\tau=0} = \boldsymbol{y}_0$ and $\boldsymbol{y}(t_f)\big|_{\tau=0} = \boldsymbol{y}_f$.

**Remark 2:** Equations (7)-(9) render a asymptotically converge to the extremum of Problem 1, while the limited time convergence may be achieved by

$$\frac{\delta \boldsymbol{y}}{\delta \tau} = -\boldsymbol{K} \cdot \mathrm{sign}\left(F_y - \frac{\mathrm{d}}{\mathrm{d}t}(F_{\dot{y}})\right), \quad t \in (t_0, t_f) \tag{12}$$

$$\frac{\delta \boldsymbol{y}(t_0)}{\delta \tau} = \boldsymbol{K} \cdot \mathrm{sign}\left(F_{\dot{y}}(t_0)\right) \tag{13}$$

$$\frac{\delta \boldsymbol{y}(t_f)}{\delta \tau} = -\boldsymbol{K} \cdot \mathrm{sign}\left(F_{\dot{y}}(t_f)\right) \tag{14}$$

where $\mathrm{sign}(\boldsymbol{a}) = \begin{bmatrix} \mathrm{sign}(a_1) \\ \mathrm{sign}(a_2) \\ ... \\ \mathrm{sign}(a_n) \end{bmatrix}$ is the sign function.

To verify the method, we consider an example from [14].

**Example 1:** Solve the scalar variable $y$, which minimizes the following functional

$$J = \int_0^\pi \left( \dot{y}^2 - 2y\cos(t) \right) \mathrm{d}t$$

with prescribed boundary conditions $y(0) = 0$ and $y(\pi) = 0$.

In solving this example with the VEM, $y(t)$ was directly discretized with 101 uniformly-spaced points as $y_i$ ($i = 1, 2, ..., 101$), and the initial guess was given by $\tilde{y}(t) = 0$. Thus a dynamic system with 101 states was obtained through Eqs. (7)-(9) with one-dimensional matrix $K = 0.1$. The calculus-of-variations problem was transformed to an IVP. The Ordinary Differential Equation (ODE) integrator "ode45" in Matlab was employed to solve the IVP and the finite difference method was used to compute the first-order and second-order derivatives at the discretization points. The analytic solution obtained through solving the BVP is

$$\hat{y} = \cos(t) + (2/\pi)t - 1$$

Fig. 2 compares the solutions, and shows that the integration results approach the optimal solution quickly. At $\tau = 6$s, it is hard to distinguish the numerical solution from the analytic. Fig. 3 plots the profile of functional value against the variation time. It monotonously declines and approaches the minimum rapidly.

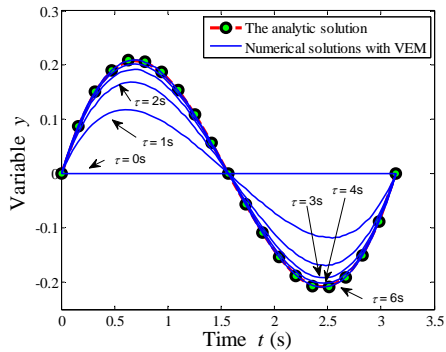

Fig. 2. The evolving of numerical solution of *y* to the analytic solution.

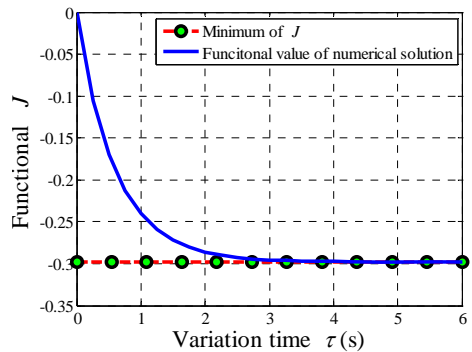

Fig.3 The approach to the minimum of functional.

## III. OPTIMAL CONTROL PROBLEMS WITH DYNAMIC CONSTRAINT

In this section, we consider the optimal control problem with dynamic equation constraint.

**Problem 2:** Consider performance index of Bolza form

$$J = \varphi(\boldsymbol{x}(t_f),t_f) + \int_{t_0}^{t_f} L(\boldsymbol{x}(t),\boldsymbol{u}(t),t)\mathrm{d}t \tag{15}$$

subject to the dynamic equation

$$\dot{\boldsymbol{x}} = \boldsymbol{f}(\boldsymbol{x},\boldsymbol{u},t) \tag{16}$$

where $t \in \mathbb{R}$ is time. $\boldsymbol{x} \in \mathbb{R}^n$ is state vector and its elements belong to $C^2[t_0,t_f]$. $\boldsymbol{u} \in \mathbb{R}^m$ is control vector and its elements belong to $C^1[t_0,t_f]$. The function $L: \mathbb{R}^n \times \mathbb{R}^m \times \mathbb{R} \to \mathbb{R}$ and its first-order and second-order partial derivatives are continuous with respect to $\boldsymbol{x}$, $\boldsymbol{u}$ and $t$. The function $\varphi: \mathbb{R}^m \times \mathbb{R} \to \mathbb{R}$ and its first-order and second-order partial derivatives are continuous with respect to $\boldsymbol{x}$ and $t$. The vector function $\boldsymbol{f}: \mathbb{R}^n \times \mathbb{R}^m \times \mathbb{R} \to \mathbb{R}^n$ and its first-order and second-order partial derivatives are continuous and Lipschitz in $\boldsymbol{x}$, $\boldsymbol{u}$ and $t$. The initial time $t_0$ is fixed and the terminal time $t_f$ is free. The initial boundary conditions are prescribed as

$$\boldsymbol{x}(t_0) = \boldsymbol{x}_0 \tag{17}$$

and the terminal states are free. Find the extremum $(\hat{\boldsymbol{x}},\hat{\boldsymbol{u}})$ that minimizes $J$, i.e.

$$(\hat{\boldsymbol{x}},\hat{\boldsymbol{u}}) = \arg\min(J) \tag{18}$$

It is well known that using the adjoining method [13], this problem may be reformulated as an unconstrained augmented functional to be

$$\bar{J} = \varphi(\boldsymbol{x}(t_f),t_f) + \int_{t_0}^{t_f} \left(L + \boldsymbol{\lambda}^{\mathrm{T}}(\boldsymbol{f} - \dot{\boldsymbol{x}})\right)\mathrm{d}t \tag{19}$$

where $\boldsymbol{\lambda}$ is the costate vector. From this augmented performance index, the first-order optimality condition may be obtained as

$$\dot{\boldsymbol{x}} - H_{\boldsymbol{\lambda}} = \boldsymbol{0} \tag{20}$$

$$\dot{\boldsymbol{\lambda}} + H_{\boldsymbol{x}} = \boldsymbol{0} \tag{21}$$

$$H_{\boldsymbol{u}} = \boldsymbol{0} \tag{22}$$

with transversality conditions

$$\boldsymbol{\lambda}(t_f) = \varphi_{\boldsymbol{x}}(t_f) \tag{23}$$

$$H(t_f) + \varphi_{t_f} = 0 \tag{24}$$

where $H = L + \lambda^T f$ is the Hamiltonian. Although the optimality conditions of Problem 2 may be deduced from Eq. (19), the extremum cannot be computed from it. This is because extremums in Problem 2 are tuned to saddle points in the augmented functional [15], and applying the VEM directly will not produce the right solution. Practically, we may construct an equivalent unconstrained functional problem that has the same extremum as Problem 2, that is

**Problem 3:** Consider the following unconstrained functional

$$J_1 = \left(H(t_f) + \varphi_{t_f}\right)^2 + \int_{t_0}^{t_f} \left\{ (\dot{x} - H_\lambda)^T (\dot{x} - H_\lambda) + (\dot{\lambda} + H_x)^T (\dot{\lambda} + H_x) + H_u^T H_u \right\} dt \tag{25}$$

where $x$, $\lambda \in \mathbb{R}^n$ and their elements belong to $C^2[t_0, t_f]$, $u \in \mathbb{R}^m$ and its elements belong to $C^1[t_0, t_f]$. The Hamiltonian $H$ and its first-order and second-order partial derivatives, with respect to $x$, $\lambda$, $u$ and $t$, are continuous. The initial time $t_0$ is fixed and the terminal time $t_f$ is free. The boundary conditions are prescribed as

$$x(t_0) = x_0 \tag{26}$$

$$\lambda(t_f) = \varphi_x(t_f) \tag{27}$$

Find the extremum $(\hat{x}, \hat{\lambda}, \hat{u})$ that minimizes $J_1$, i.e.

$$(\hat{x}, \hat{\lambda}, \hat{u}) = \arg\min(J_1) \tag{28}$$

It is readily to find that the extremum of Problem 2 is also the extremum of Problem 3. Especially, Problem 3 is a convex functional optimization problem, and its solution is the one that satisfies Eqs. (20)-(22). Replacing the function and variables in Eq. (1) respectively with

$$F = (\dot{x} - H_\lambda)^T (\dot{x} - H_\lambda) + (\dot{\lambda} + H_x)^T (\dot{\lambda} + H_x) + H_u^T H_u \tag{29}$$

$$y = \begin{bmatrix} x \\ \lambda \\ u \end{bmatrix} \tag{30}$$

According to the VEM and with extra consideration on the free terminal time, we may deduce the variation dynamic evolving equations as

$$\frac{\delta y}{\delta \tau} = -2Kr, \quad t \in (t_0, t_f) \tag{31}$$

$$\frac{\delta y(t_0)}{\delta \tau} = 2K \begin{bmatrix} \mathbf{0} \\ (\dot{\lambda} + H_x) \\ r_u \end{bmatrix}\bigg|_{t_0} \quad (32)$$

$$\frac{\delta y_f}{\delta \tau} = -2K \begin{bmatrix} (H + \varphi_{t_f})(H_x + \varphi_{xt_f}) + (\dot{x} - H_\lambda) \\ \mathbf{0} \\ (H + \varphi_{t_f}) \cdot H_u \end{bmatrix}\bigg|_{t_f} \quad (33)$$

$$\frac{\delta t_f}{\delta \tau} = -k_{t_f}\left(2(H+\varphi_{t_f})(H_t + \varphi_{t_f t_f}) + H_x^T H_x + H_\lambda^T H_\lambda + H_u^T H_u - \dot{x}^T \dot{x} - \dot{\lambda}^T \dot{\lambda}\right)\bigg|_{t_f} \quad (34)$$

with initial value $x(t_0)|_{\tau=0} = x_0$ and $\lambda(t_f)|_{\tau=0} = \varphi_x(t_f)$, where

$$r = \begin{bmatrix} r_x \\ r_\lambda \\ r_u \end{bmatrix} = H_{yy} v + M\dot{y} + \begin{bmatrix} f_t \\ -H_{xt} \\ \mathbf{0} \end{bmatrix} - \begin{bmatrix} \ddot{x} \\ \ddot{\lambda} \\ \mathbf{0} \end{bmatrix} \quad (35)$$

$H_{yy} = \begin{bmatrix} H_{xx} & f_x^T & H_{xu} \\ f_x & 0 & f_u \\ H_{ux} & f_u^T & H_{uu} \end{bmatrix}$ is the Hessian matrix, $v = \begin{bmatrix} (H_x + \dot{\lambda}) \\ (f - \dot{x}) \\ H_u \end{bmatrix}$ is the optimality vector, the matrix $M$ is

$M = \begin{bmatrix} f_x & 0 & f_u \\ -H_{xx} & -f_x^T & -H_{xu} \\ 0 & 0 & 0 \end{bmatrix}$, $\ddot{x} = \frac{d^2 x}{dt^2}$ and $\ddot{\lambda} = \frac{d^2 \lambda}{dt^2}$. $\frac{\delta y_f}{\delta \tau} = \frac{\delta y(t_f)}{\delta \tau} + \dot{y}\frac{\delta t_f}{\delta \tau}$ is the derivative of variation in the terminal variable with respect to $\tau$. $K = \text{diag}(k_1, k_2, ..., k_{2n+m})$ is the positive diagonal matrix and $k_{t_f}$ is a positive constant.

**Theorem 2:** Solving the IVP defined by Eqs. (31)-(34) from the VEM, when $\tau \to +\infty$, we have $J_1 \to 0$ and $(x, \lambda, u)$ will satisfy the optimality conditions of Problem 2.

Proof: Differentiating Eq. (25) with respect to $\tau$ and substituting Eqs. (31)-(34) in, we have $\frac{\delta J_1}{\delta \tau} \leq 0$. The convex functional $J_1$ will decrease until $J_1 = 0$, which occurs when $\tau \to +\infty$. Because the extremum of the unconstrained functional defined in Problem 3 satisfies the optimality conditions of Problem 2, when $J_1$ reaches the minimum, i.e., $J_1 = 0$, this determines the optimality conditions of Problem 2. ∎

**Remark 3:** If in Problem 2 the terminal time $t_f$ is fixed, then the equivalent unconstrained functional may be

$$J_1 = \int_{t_0}^{t_f} \left\{(\dot{x} - H_\lambda)^T(\dot{x} - H_\lambda) + (\dot{\lambda} + H_x)^T(\dot{\lambda} + H_x) + H_u^T H_u\right\} dt \quad (36)$$

Using the VEM, the variation dynamic evolving equations deduced is similar as the free $t_f$ case except that Eq. (34) is not applicable and Eq. (33) is reformulated as

$$\frac{\delta y_f}{\delta \tau} = -2K \begin{bmatrix} (\dot{x} - H_\lambda) \\ 0 \\ r_u \end{bmatrix}_{t_f} \tag{37}$$

with the initial value $\lambda(t_f)\big|_{\tau=0} = \varphi_x(t_f)$.

**Remark 4:** If in Problem 2 the terminal boundary condition of the states is prescribed as $x(t_f) = x_f$, then using the VEM the variation dynamic evolving equations deduced is similar as the free terminal states case except Eq. (33) is reformulated as

$$\frac{\delta y_f}{\delta \tau} = -2K \begin{bmatrix} 0 \\ (H + \varphi_{t_f}) \cdot H_\lambda + (\dot{\lambda} + H_x) \\ (H + \varphi_{t_f}) \cdot H_u \end{bmatrix}_{t_f} \tag{38}$$

with the initial value $x(t_f)\big|_{\tau=0} = x_f$.

**Remark 5:** If in Problem 2 the terminal time $t_f$ is fixed and the terminal boundary condition of the states is prescribed as $x(t_f) = x_f$, then using the functional (36), the variation dynamic evolving equations deduced by the VEM is similar as the free terminal states case except Eq. (37) is reformulated as

$$\frac{\delta y_f}{\delta \tau} = -2K \begin{bmatrix} 0 \\ (\dot{\lambda} + H_x) \\ r_u \end{bmatrix}_{t_f} \tag{39}$$

with the initial value $x(t_f)\big|_{\tau=0} = x_f$.

**Remark 6:** Analogous to the Lyapunov function in the nonlinear dynamic system, the functional (25) itself is a reasonable index to measure the "distance" to the extremum. With the VEM, it is globally monotonously decreases and this ensures that $(x, \lambda, u)$ will approach the extremum, i.e., $(\hat{x}, \hat{\lambda}, \hat{u})$, over time.

An example from [16] is now considered.

**Example 2**: Consider the following dynamic system

$$\dot{x} = Ax + bu$$

where $x = \begin{bmatrix} x_1 \\ x_2 \end{bmatrix}$, $A = \begin{bmatrix} 0 & 1 \\ 0 & 0 \end{bmatrix}$, $b = \begin{bmatrix} 0 \\ 1 \end{bmatrix}$. Find the solution that minimizes the performance index

$$J = \frac{1}{2} \int_{t_0}^{t_f} u^2 \mathrm{d}t$$

with the boundary conditions

$$x(t_0) = \begin{bmatrix} 1 \\ 1 \end{bmatrix}, \quad x(t_f) = \begin{bmatrix} 0 \\ 0 \end{bmatrix}$$

where the initial time $t_0 = 0$ and the terminal time $t_f = 2$ are fixed.

As the principle, the equivalent unconstrained functional was constructed as Eq. (36) where $H = \frac{1}{2}u^2 + \lambda^\mathrm{T} Ax + \lambda^\mathrm{T} bu$, $H_\lambda = Ax + bu$, $H_x = A^\mathrm{T} \lambda$ and $H_u = u + \lambda^\mathrm{T} b$. We also discretized the time horizon $[t_0, t_f]$ uniformly, with 41 points. The initial guess was given by $\tilde{x}(t) = \begin{bmatrix} 1 - 0.5t \\ 1 - 0.5t \end{bmatrix}$, $\tilde{\lambda}(t) = \begin{bmatrix} 0 \\ 0 \end{bmatrix}$ and $\tilde{u}(t) = 0$. Through Eqs. (31), (32) and (39), with parameters $k_i \in K$ and $k_{t_f}$ equaling 1, we obtained a large IVP with 205 states. Since we did not specially scale the problem, we employed a stiff ODE integration method, "ode15s" in Matlab, for the numerical integration. The analytic solution by solving the BVP is

$$\begin{aligned}
\hat{x}_1 &= 0.5t^3 - 1.75t^2 + t + 1 \\
\hat{x}_2 &= 1.5t^2 - 3.5t + 1 \\
\hat{\lambda}_1 &= 3 \\
\hat{\lambda}_2 &= -3t + 3.5 \\
\hat{u} &= 3t - 3.5
\end{aligned}$$

Figs. 4 and 5 show the evolving process of $x_1$ and $\lambda_1$ solutions to the optimal respectively, and at $\tau = 300$s, the numerical solutions are indistinguishable from the analytic. Fig. 6 plots the numerical solutions of $x_2$, $\lambda_2$, and $u$ at $\tau = 300$s. They are almost identical with the analytic and this shows the effectiveness of the VEM.

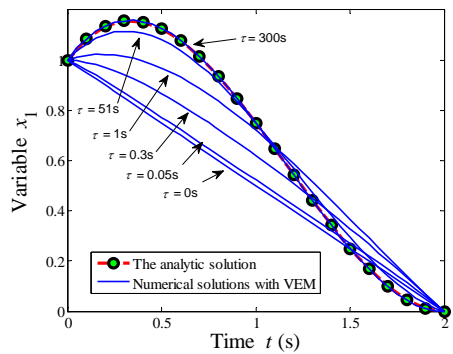

Fig. 4 The evolving of numerical solution of $x_1$ to the analytic solution.

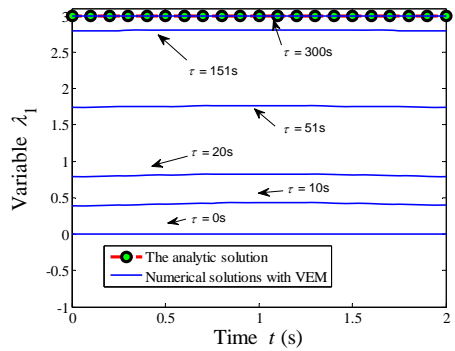

Fig. 5 The evolving of numerical solution of $\lambda_1$ to the analytic solution.

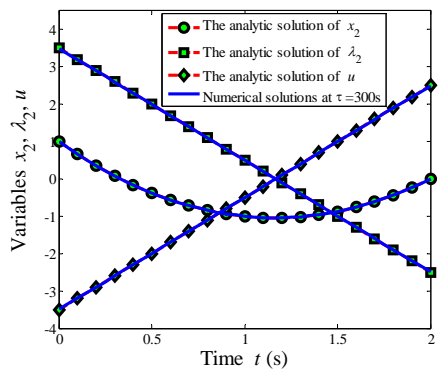

Fig. 5 The numerical solutions of $x_2$, $\lambda_2$, and $u$ with VEM at $\tau = 300$s .

Now we consider a nonlinear example with free terminal time $t_f$, the Brachistochrone problem [17], which describe the motion curve of the fastest descending.

**Example 3**: Consider the following dynamic system

$$\dot{x} = f(x,u)$$

where $x = \begin{bmatrix} x \\ y \\ V \end{bmatrix}$, $f = \begin{bmatrix} V\sin(u) \\ -V\cos(u) \\ g\cos(u) \end{bmatrix}$, $g = 10$ is the gravity constant. Find the solution that minimizes the performance index

$$J = t_f$$

with the boundary conditions

$$\begin{bmatrix} x \\ y \\ V \end{bmatrix}\bigg|_{t_0=0} = \begin{bmatrix} 0 \\ 0 \\ 0 \end{bmatrix}, \begin{bmatrix} x \\ y \end{bmatrix}\bigg|_{t_f} = \begin{bmatrix} 2 \\ -2 \end{bmatrix}$$

This example has fixed terminal position boundary conditions and free terminal velocity $V(t_f)$. Thus, the boundary state variation dynamic equations are synthesized accordingly between Eqs. (33) and (38). An initial guess of $t_f\big|_{\tau=0} = 1$ s was used and the time horizon was again discretized uniformly, with 101 points. The initial guess was given by $\tilde{x}(t) = \begin{bmatrix} 1-t \\ 1-t \\ 0 \end{bmatrix}$, $\tilde{\lambda}(t) = \begin{bmatrix} 0 \\ 0 \\ 0 \end{bmatrix}$, and $\tilde{u}(t) = 0$. We also used "ode15s" to solve the large IVP. Here we computed the optimal solution with GPOPS-II [18], a Radau PS method based OCP solver. Fig. 7 gives the states curve in $xy$ position plane, showing that the numerical result approaches the optimal solution over time. In Fig. 8 the terminal time profile against the variation time $\tau$ is plotted. The result of $t_f$ increases rapidly at first and then gradually decreases to the minimum decline time, and it almost keeps unchanged after $\tau = 170$s. At $\tau = 400$s, we compute that $t_f = 0.8166$s, very close to the optimal result of 0.8165s.

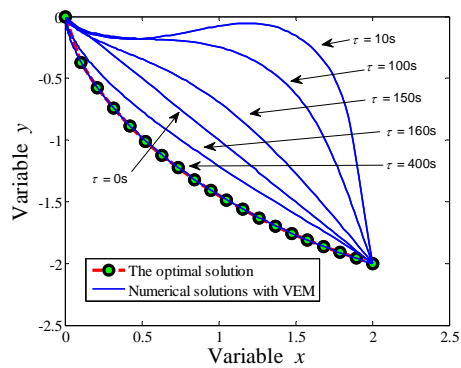

Fig. 7  The evolving of numerical solution in $xy$ position plane to the optimal solution.

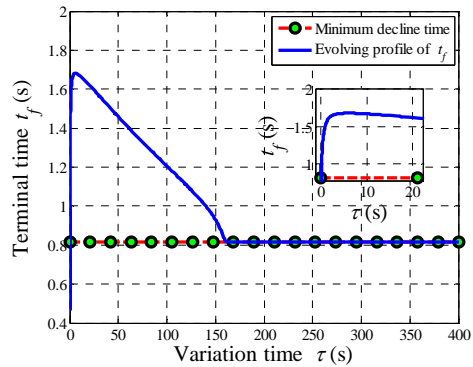

Fig.8  The evolving profile of $t_f$ to the minimum decline time.

**IV. FURTHER COMMENTS**

To facilitate the propagation of the discovery and acknowledge the authors' work, if not improper, the equations (31)-(34) may be called the ZS equation. Actually, without considering the boundary conditions at $t_0$ and $t_f$, the ZS equation may be presented as a Partial Differential Equation (PDE) by replacing the variation operation "$\delta$" with the partial differential operator "$\partial$" to be

$$\frac{\partial}{\partial \tau}\begin{bmatrix} x \\ \lambda \\ u \end{bmatrix} = -2K\left( H_{yy} \begin{bmatrix} \left(H_x + \frac{\partial \lambda}{\partial t}\right) \\ \left(f - \frac{\partial x}{\partial t}\right) \\ H_u \end{bmatrix} - \frac{\partial}{\partial t}\begin{bmatrix} \left(\frac{\partial x}{\partial t} - f\right) \\ \left(\frac{\partial \lambda}{\partial t} + H_x\right) \\ 0 \end{bmatrix}\right) \quad (40)$$

Consider in this way, the former examples are actually solved by the well-known semi-discrete method in the field of PDE numerical calculation [19]. For Eq. (40), recall Fig. 1, the initial boundary conditions of $x(t,\tau)$, $\lambda(t,\tau)$ and $u(t,\tau)$ at $\tau = 0$ may be arbitrary and their value at $\tau = +\infty$ is the extremal solution of the OCP. Since the boundary value at $\tau = +\infty$ is the truly demanded, efficient methods may be developed for computing this PDE.

For the state- and control-constrained OCPs, the strategy developed above is not generally applicable, because the analogous equivalent functional is not available when complex path constraints are involved. Further studies along that thread may require employment of techniques such as the Karush–Kuhn–Tucker (KKT) variables or the slack variables [20]. However, for a particular class of OCP, the time-optimal control problems with control constraint, an effective alternative may be developed and this will appear in a forthcoming paper.

Actually, any discrete iteration method may have its continuous-time evolving counterpart theoretically. For example, from the same principle, the continuous-time dynamic equation may also be derived for the solutions of parameter optimization problems. Since the optimization procedure is driven by the "engine" of ODE integration methods, this may motivate us to develop efficient and easy-to-use optimization framework.

## V. Conclusion

In this paper, a novel method, originating from the stable dynamics evolution, is proposed for the optimal control solution. By using the variation motion, the Variation Evolving Method (VEM) transforms the Optimal Control Problem (OCP) to the Initial-value Problem (IVP), thus avoiding the complexity of the two-point Boundary Value Problem (BVP). Especially, because the reformulated functional, which measures the "distance" to the optimal solution, decreases globally, the solution is ensured to approach the optimal in an asymptotically way. The VEM also synthesizes the direct and indirect methods, but from a new standpoint.

Compared with the discrete iteration methods, daunting task of searching reasonable step size and annoying oscillation phenomenon around the extremum, as occurs in the discrete gradient method, are eliminated and the mature Ordinary Differential Equation (ODE) integration methods may be employed to solve the OCPs.